# Fluctuating diffusivity of RNA-protein particles: Analogy with thermodynamics

**Yuichi Itto** [1,2]

[1] ICP, Universität Stuttgart, 70569 Stuttgart, Germany

[2] Science Division, Center for General Education, Aichi Institute of Technology,

Aichi 470-0392, Japan

**Abstract:**   A formal analogy of fluctuating diffusivity to thermodynamics is discussed for messenger RNA molecules fluorescently fused to a protein in living cells. Regarding the average value of the fluctuating diffusivity of such RNA-protein particles as the analog of the internal energy, the analogs of the quantity of heat and work are identified. The Clausius-like inequality is shown to hold for the entropy associated with diffusivity fluctuations, which plays a role analogous to the thermodynamic entropy, and the analog of the quantity of heat. The change of the statistical fluctuation distribution is also examined from a geometric perspective. The present discussions may contribute to a deeper understanding of the fluctuating diffusivity in view of the laws of thermodynamics.

## 1. Introduction

There are growing experimental observations showing exotic physical properties of messenger RNA molecules in living cells. A recent experimental study in Ref. [1] has offered one such example. The RNA molecules, each of which is fluorescently labeled with a protein, exhibit a heterogeneous diffusion phenomenon with fluctuating diffusivity for two different types of cell: *Escherichia coli* cell (i.e., a bacterium) and *Saccharomyces cerevisiae* cell (i.e., a yeast). For individual trajectories of such RNA-protein particles uniformly distributed over cytoplasm of each cell, analysis of the mean square displacement, which behaves for elapsed time, $t$, as

$$\overline{x^2} \sim D t^\alpha, \tag{1}$$

has revealed that the diffusivity, $D$, i.e., the diffusion coefficient in units of $[\mu\mathrm{m}^2/\mathrm{s}^\alpha]$, fluctuates in a wide range, whereas the diffusion exponent, $\alpha$, is approximately constant, taking a certain positive value less than unity. The latter reflects viscoelastic nature of the cytoplasm [1] (see also, for example, Refs. [2-4] for relevant experimental works) and highlights the phenomenon referred to as anomalous diffusion [5], which is of great interest for various disciplines in the literature [6,7], showing subdiffusivity, i.e., $0 < \alpha < 1$.

As can be seen in Fig. 3 in Ref. [1], the diffusivity obeys the following exponential law:

$$P_0(D) \sim \exp(-D/D_0), \tag{2}$$

where $D_0$ is the average value of $D \in [0, \infty)$ and yields a typical value of the diffusivity. There, it is appreciated that the different data in the two cell types follow this law, showing robustness of the exponential diffusivity fluctuations. The distribution in Eq. (2) has played a key role for obtaining the non-Gaussian distribution of the displacements of the RNA-protein particles [1]. (Such a role can also be found in recent works in Refs. [8,9], for example.)

In spite of its simple form, the origin of the exponential diffusivity fluctuations remains unclear. However, the following idea has been suggested [1]: the distribution in Eq. (2) is the *maximal entropy* distribution. To accomplish this, a maximum-entropy-principle approach has been developed in a recent work in Ref. [10]. Its basic observation is as follows. The cytoplasm is regarded as a medium consisting of many local blocks (or, regions), a typical size of which has also been estimated based on the experimental data. In each of these local blocks, the diffusivity in Eq. (1) slowly varies on a time scale much larger than that of dynamics of the RNA-protein particles. The quantity, $S$, is then introduced as a measure of uncertainty about local diffusivity fluctuations over the medium, which turns out to take the form of the Shannon entropy [11] given by

$$S[P] = -\int dD P(D) \ln P(D) \qquad (3)$$

with $P(D) dD$ being the probability of finding the diffusivity in the interval $[D, D + dD]$. Due to the large separation of time scales, it is considered that the fluctuation distribution to be observed may maximize this entropy. Together with the constraint on the normalization condition, $\int dD P(D) = 1$, maximization of the entropy with respect to the fluctuation distribution under the constraint on the expectation value of the diffusivity, $\int dD P(D) D = \overline{D}$, is found to give the following exponential distribution, $\hat{P}(D) \propto \exp(-\lambda D)$ with $\lambda$ being a positive Lagrange multiplier associated with the constraint in the latter, showing that it becomes the distribution in Eq. (2) after the identification $\lambda = 1/D_0$. We here mention that this exponential distribution has formally the form of the canonical distribution [12], if it is assumed that $D_0$ is proportional to the average value of temperature over the local blocks [see Eq. (11) below], and this fact turns out to constitute a key in our later discussion.

The above idea has been further supported by explicitly showing [10] that the entropy production rate becomes manifestly positive under the mechanism of the so-called "diffusing diffusivity" [13], which offers a description of time evolution of diffusivity fluctuations and leads to the exponential fluctuation distribution as a stationary solution of its evolution equation (see, e.g., Ref. [14] for a recent development, where emergence of correlation time characterizing diffusivity dynamics has been discussed). As shown in Ref. [1], this mechanism combined with the approach of fractional Brownian motion [15] modelling the subdiffusion of the RNA-protein particles yields the non-Gaussian displacement distribution observed in the experiment. (In Ref. [16], it has been found that there exists the lower bound on the rate suppressing the entropy production, which is independent of time.)

Now, a comment, which motivates our present work, has also been made on an analogy with the thermodynamic relation concerning temperature [10]. Let us consider the thermodynamic-like situation in such a way that $D_0$ slowly changes, the time scale of which should be much larger than that of variation of diffusivity fluctuations mentioned above. It is shown, for the entropy in Eq. (3) with the distribution in Eq. (2), that $\partial S/\partial D_0 = 1/D_0$. It is also assumed that the diffusivity is proportional to temperature in the local block [see also the discussion after Eq. (7) below], like in the Einstein relation [17], and temperature slowly fluctuates depending on the blocks, (the values of which are denoted by $T_i$'s discussed below), i.e., the medium is in nonequilibrium-stationary-state-like situation. (In fact, such local temperature fluctuations are expected to be realized, see Ref. [18].) Under this, the following relation then holds:

$$\frac{\partial \tilde{S}}{\partial D_0} = \frac{1}{T}, \tag{4}$$

provided that

$$\tilde{S} = cS, \quad D_0 = cT, \tag{5}$$

where $T$ denotes the average value of temperature over the local blocks and $c$ is a positive quantity characterizing mobility of the RNA-protein particles. [It is noticed that maximization of $\tilde{S}$ also leads to the distribution in Eq. (2) after the redefinition of the Lagrange multipliers.] The derivative appearing in Eq. (4) indicates that the volume of the local block is kept unchanged, implying that $c$ is fixed, which is discussed in our later discussion. At the statistical level, the relation therefore has an analogy with the thermodynamic relation concerning temperature, *if $\tilde{S}$ and $D_0$ are identified with the analogs of the "thermodynamic entropy" and the "internal energy", respectively*.

In this paper, we study a formal analogy of the fluctuating diffusivity to thermodynamics for the RNA-protein particles in cytoplasm of *Escherichia coli* cell as well as *Saccharomyces cerevisiae* cell. Regarding the average value of the fluctuating diffusivity as the analog of the internal energy, we identify the analog of the quantity of heat as well as that of work. We also show that the analog of the Clausius inequality holds for the entropy associated with diffusivity fluctuations, which is analogous to the thermodynamic entropy, and the analog of the quantity of heat. The change of diffusivity fluctuation distribution for realizing these analogs is also discussed from a geometric perspective. Thus, the present discussions may give a step toward understanding the fluctuating diffusivity from the viewpoint of the laws of thermodynamics.

**2. Analogs of the quantity of heat and work**

Consider the medium in a certain state with a set of different diffusivities, $\{D_i\}_i$, where $D_i$ denotes the $i$ th value of the diffusivity and slowly varies. Here and hereafter, we purposely develop our discussion in the discrete case of the diffusivity. Let us regard the average value of the diffusivity with respect to some fluctuation distribution $P(D_i)$ to be observed as the analog of the internal energy:

$$U_D = \sum_i D_i P(D_i). \tag{6}$$

Due to the slow variation of the fluctuations, it is assumed that $P(D_i)$ deviates from the exponential distribution in Eq. (2) slightly, in general. In the thermodynamic-like situation, the medium is considered to be found in the state with the local diffusivity fluctuations with a certain statistical fluctuation, and these states are distinct each other in the sense that the local property of diffusivity fluctuations in a given state differs infinitesimally from that in the other states. Along a process connecting two such states, the change of $U_D$ is given by

$$\delta U_D = \sum_i D_i \delta P(D_i) + \sum_i P(D_i) \delta D_i, \qquad (7)$$

where $\delta P(D_i)$ stands for the change of the statistical form of the fluctuation distribution, whereas $\delta D_i$ describes the change of the diffusivity due to external influence. In the case when the statistical fluctuation takes the exponential form in Eq. (2), the medium is supposed to be in the state analogous to the "equilibrium state".

We here discuss the following points. The experimental results in Ref. [1] have supported the approach of fractional Brownian motion [15] as an underlying stochastic process for modelling subdiffusion of the RNA-protein particles. Then, in Refs. [19,20], it has been shown, for the subdiffusive behavior of random walkers such as the RNA-protein particles in *Escherichia coli* cells, that the mean square displacement of the walkers is proportional to temperature for large elapsed time, based on a generalized Langevin equation characterizing viscoelastic nature of the cytoplasm, which is known to offer the subdiffusivity equivalent to fractional Brownian motion, see, e.g., Ref. [21]: the proportionality factor includes the friction constant depending on the diffusion exponent, which reflects the viscoelasticity linked to elastic elements such as cytoskeletal filaments (see also, e.g., Ref. [22] for a similar approach in this context). It is of interest to experimentally examine if these features are observed.

It may be worth to point out that such features have also been discussed in a recent work in Ref. [23] for DNA-binding proteins in *Escherichia coli* cells.

Therefore, like in the Einstein relation [17], considering a set of different temperatures, $\{T_i\}_i$, with $T_i$ being the $i$ th value of temperature in the local blocks, we shall assume the following relation $D_i = cT_i$, where $c$ is supposed not to drastically alter over the local blocks, recalling that $\alpha$ in Eq. (1) is approximately constant and the proportionality factor mentioned above depends on the diffusion exponent. In the case when $P(D_i) = P_0(D_i)$, the average value of $T_i$ in its continuum limit is given by $T$ in Eq. (5). Under these, both $\delta P(D_i)$ and $\delta D_i$ may be realized by the change of temperature and expansion/compression of the cell. In fact, it has experimentally been observed in a recent work in Ref. [24] (see also references therein) that cytoplasmic particles exhibit the subdiffusive behavior in *Escherichia coli* cells. It has been then found that the average value of the diffusivity of such particles decreases under compression of the cells. There, it seems natural to consider that this compression process yields a mechanical external influence, while temperature, which the cells are subject to, remains unchanged. In such a situation in the present context, from $D_i = cT_i$, $\delta D_i$ is expected to come from the change of $c$, implying that $c$ plays a role analogous to external parameter.

Since we are considering that the average value of the diffusivity is the analog of the internal energy, it is natural to identify the analog of work as

$$\delta' W_D = -\sum_i P(D_i)\delta D_i, \tag{8}$$

and the analog of the quantity of heat is identified as

$$\delta' Q_D = \sum_i D_i \delta P(D_i), \tag{9}$$

accordingly. Therefore, we obtain the analog of the first law of thermodynamics [12]:

$$\delta' Q_D = \delta U_D + \delta' W_D. \tag{10}$$

### 3. Dependences on temperature and the analog of external parameter

Based on the above discussions, let us examine how the analogs of the quantity of heat and work depend on temperature and the analog of external parameter for the exponential fluctuation distribution $P_0(D_i)$. To do so, it may be useful to write it as

$$P_0(D_i) = \frac{1}{Z}\exp(-D_i/D_0), \quad Z = \sum_i \exp(-D_i/D_0), \tag{11}$$

with $D_0 = cT$ in Eq. (5), which formally takes the form of the canonical distribution [12]. It is noticed that the value of $D_i/D_0$ does not depend on $c$. From this, in Eq. (7) with Eq. (11), it is understood that the first term on the right-hand side is associated with $\delta P(D_i)$ due to the change of $T$, whereas the second term is related to $\delta D_i$ with the statistical form being kept unchanged, suggesting that its origin comes from the change of $c$ only.

Therefore, from Eqs. (8) and (9), we have

$$\delta' Q_D = \frac{\langle (D_i - \langle D_i \rangle)^2 \rangle}{cT^2} \delta T \tag{12}$$

and

$$\delta' W_D = -\frac{\partial \langle D_i \rangle}{\partial c} \delta c, \tag{13}$$

where $\delta T$ and $\delta c$ describe the changes of $T$ and $c$, respectively, and the angle brackets denote the average with respect to the fluctuation distribution in Eq. (11). It may be of interest to see that the analog of the quantity of heat in Eq. (12) is described by the variance of the diffusivity. Equivalently, with $Z$, Eqs. (12) and (13) are expressed as

$$\delta' Q_D = c \frac{\partial}{\partial T}\left(T^2 \frac{\partial \ln Z}{\partial T}\right) \delta T \tag{14}$$

and

$$\delta' W_D = -T^2 \frac{\partial \ln Z}{\partial T} \delta c. \tag{15}$$

## 4. Analog of the Clausius inequality

In this section, we establish the analog of the second law of thermodynamics. (Our discussion is based on a basic observation in Ref. [25], where robustness of this law has been studied for a generalized entropic measure in the context of complex systems in nonequilibrium stationary states.) For the above-mentioned infinitesimal process, we first evaluate the change of the entropy, $S[P] = -\sum_i P(D_i) \ln P(D_i)$, for the exponential fluctuation distribution. Under the normalization condition on $P(D_i)$, the entropy change is given by

$$\delta S = -\sum_i \left( \ln P(D_i) \right) \delta P(D_i)$$
$$= \frac{\delta' Q_D}{D_0}, \tag{16}$$

where Eq. (2) (in the discrete case) has been used at the second equality. In the case when the quantity $c$ is fixed, which implies that the volume of the local block remains unchanged consistently with the situation in Eq. (4), using the relations in Eq. (5), we have

$$\delta \tilde{S} = \frac{\delta' Q_D}{T}, \tag{17}$$

where it is understood that $\tilde{S}$ is analogous to the thermodynamic entropy. This also justifies the identifications in Eqs. (8) and (9).

Next, our interest is in the entropy change in the case when the fluctuation distribution differs from the exponential one in Eq. (2). Consider the two different distributions of diffusivity fluctuations: one is the exponential fluctuation distribution $P_0(D_i)$, and the other is the fluctuation distribution $P(D_i)$. Then, we quantify the difference between them by employing the Kullback-Leibler relative entropy [26] given by

$$K[P \| P_0] = \sum_i P(D_i) \ln \frac{P(D_i)}{P_0(D_i)}, \qquad (18)$$

which is positive semidefinite and vanishes if and only if $P(D_i) = P_0(D_i)$. In the present infinitesimal process, let us write $\delta P(D_i)$ as follows:

$$\delta P(D_i) = \{\gamma P^*(D_i) + (1-\gamma) P(D_i)\} - P(D_i), \qquad (19)$$

where $P^*(D_i)$ denotes some fluctuation distribution satisfying

$$K[P^* \| P_0] \leq K[P \| P_0], \qquad (20)$$

and $\gamma$ is a constant in the range $0 < \gamma < 1$. For our subsequent discussion, the convexity of the relative entropy in Eq. (18) is crucial, which is given, in terms of these fluctuation distributions, as follows:

$$K[\gamma P^* + (1-\gamma)P \| P_0] \leq \gamma K[P^* \| P_0] + (1-\gamma)K[P \| P_0]. \tag{21}$$

With Eq. (19), let us evaluate the change of the relative entropy given by

$$\delta_P K[P \| P_0] = K[\gamma P^* + (1-\gamma)P \| P_0] - K[P \| P_0], \tag{22}$$

where $\delta_P$ stands for the change with respect to $P(D_i)$. What is important here is the fact that this change turns out to be not positive. In fact, from Eqs. (20)-(22), we have

$$\delta_P K[P \| P_0] \leq \gamma \{K[P^* \| P_0] - K[P \| P_0]\}$$

$$\leq 0. \tag{23}$$

The change in Eq. (22) itself is then calculated to be

$$\delta_P K[P \| P_0] = -\delta S[P] + \frac{\delta' Q_D}{D_0}, \tag{24}$$

where the normalization condition on the fluctuation distribution has been used. Consequently, in the case when $c$ is fixed, from Eqs. (5), (23), and (24), we obtain the analog of the Clausius inequality:

$$\delta \tilde{S} \geq \frac{\delta' Q_D}{T}. \tag{25}$$

## 5. Geometric perspective of the change of statistical fluctuation

We further discuss about the change in Eq. (19), which enables us to realize it based on a geometric perspective. For it, we examine Eq. (20) up to the second order of $\delta P(D_i)$, which leads to

$$\frac{1}{\gamma}\sum_i \left( \ln \frac{P(D_i)}{P_0(D_i)} \right) \delta P(D_i) + \frac{1}{2\gamma^2} \sum_i \frac{(\delta P(D_i))^2}{P(D_i)} \leq 0. \tag{26}$$

This can be then recast in the following inequality:

$$\sum_i \frac{(\delta P(D_i) - h_i)^2}{a_i^2} \leq 1, \tag{27}$$

where $h_i$ and $a_i$ are defined by

$$h_i = \gamma P(D_i) \ln \frac{P_0(D_i)}{P(D_i)} \tag{28}$$

and

$$a_i = \gamma \sqrt{\left[\sum_j P(D_j)\left(\ln \frac{P(D_j)}{P_0(D_j)}\right)^2\right] P(D_i)}, \qquad (29)$$

respectively. This means that regarding $\{\delta P(D_i)\}_i$ as a coordinate point in the space of the change, under the condition, $\sum_i \delta P(D_i) = 0$, the change giving rise to the analog of the Clausius inequality in Eq. (25) exists inside an ellipsoid centered at the point characterized by $\{h_i\}_i$ with the semi-axes $a_i$'s. In other words, the condition $\sum_i \delta P(D_i) = 0$ yields a hyperplane passing through the origin in the space, which intersects with the ellipsoid: this gives a set of points of the intersection inside the ellipsoid, corresponding to the change.

Therefore, these observations enable us to see a geometric perspective of the change of statistical fluctuation distribution associated with the analogy between the fluctuating diffusivity and thermodynamics, suggesting how such a change is performed.

## 6. Conclusion

We have developed a formal analogy between fluctuating diffusivity and thermodynamics for RNA-protein particles in both *Escherichia coli* cell and *Saccharomyces cerevisiae* cell. Regarding the average value of fluctuating diffusivity as the analog of the internal energy, we have identified the analogs of the quantity of heat and work. For the exponential diffusivity fluctuations, we have also examined how these analogs depend on temperature and the analog of external parameter, the change of which may be realized by expansion/compression of the cell. In analogy of the entropy associated with diffusivity fluctuations to the thermodynamic entropy, we have then established the analog of the Clausius inequality for the entropy and the analog of the quantity of heat. This analog has also been discussed from a geometric perspective, suggesting how the change of the statistical fluctuation distribution is done. Therefore, these results allow us to study the fluctuating diffusivity from the viewpoint of the laws of thermodynamics.

In Ref. [27], entropy concerned with the fluctuations of the volume of granules in soft materials has been discussed in the context of subdiffusion. It may be of interest to examine such a discussion for the RNA-protein particles, if expansion/compression of the particles is considered, where the analogy of the quantity $c$ to external parameter may be found.

**Acknowledgements**

The present work is based on the author's talk at the 45th Conference of the Middle European Cooperation in Statistical Physics (14-16 September, 2020). He would like to acknowledge the organizers of the conference for providing him with opportunity to give a talk. The work has been completed while he has stayed at the Institut für Computerphysik, Universität Stuttgart. He would like to thank the Institut für Computerphysik for the warm hospitality.